\documentstyle[12pt]{article}
\def\be{\begin{equation}}
\def\ee{\end{equation}}
\def\bea{\begin{eqnarray}}
\def\eea{\end{eqnarray}}
\def\ed{\end{document}}
\def\bd{
\begin{document}}
\def\bit{\begin{itemize}}
\def\eit{\end{itemize}}

% Greek letters

\def\sig{\sigma}
\def\Sig{\Sigma}
\def\lam{\lambda}
\def\Lam{\Lambda}
\def\Del{\Delta}
\def\del{\delta}
\def\Bg{\Bar g}
\def\hg{\hat g}
\def\k{\kappa}
\def\alf{\alpha}
\def\ga{\gamma}
\def\Ga{\Gamma}
\def\BD{\Bar D}

\def\Lix{\pounds_\xi}
\def\di{\partial}
\def\nab{\nabla}
\def\half{{\textstyle{1 \over 2}}}
\def\~{\tilde}
\def\lag{\hat{\cal L}}
\def\m{\label}
\def\l{\left}
\def\r{\right}
\def\goto{\rightarrow}
\def\Bar{\overline}
\def\const{\rm const}

%%%%%%%%%%%%%%%%%%%%%%%%%%%%%%%%%%%%%%%%%%%%%%%%%%%%%
\bd

\centerline{\bf PERTURBATIONS IN THE EINSTEIN THEORY } 
 \centerline{\bf OF GRAVITY:  COSERVED CURRENTS}

\vskip .1 in
 \centerline{Alexander N. Petrov\footnote{E-mail:
anpetrov@rol.ru}}

\centerline{Sternberg Astronomical Institute,}
\centerline{ Universitetskii
prospect 13, Moscow 119992, Russia}

\vskip .1 in
\begin{quote}
{\small \it General relativity in the form where gravitational perturbations
together with other physical fields  propagate  on an auxiliary
background is considered.  With using the Katz-Bi\v
c\'ak-Lynden-Bell technique new conserved currents, divergences
of antisymmetric tensor densities (superpotentials),  in
arbitrary curved spacetime are constructed.}
\end{quote}

\section{ A short review and setting tasks.}
Frequently,
the perturbed Einstein equations are suggested as follows.
The linear in metric perturbations terms are placed on the left hand side;
whereas all nonlinear terms are transported to the right hand side, and
together with a matter energy-momentum tensor are treated as a
total (effective) energy-momentum tensor $t^{(tot)}_{\mu\nu}$.
A such consideration was developed in a form of a theory
of a tensor field with self-interaction in a fixed
background spacetime where 
$t^{(tot)}_{\mu\nu}$ is obtained
by variation of an action with respect to a background metric 
$\Bar g^{\mu\nu}$. Sometimes it 
is called as
a {\it field formulation} of general relativity (GR)
(see, e.g., \cite{[1]}).
One of key works in the field formulation
of GR is the paper by 
Deser \cite{[2]} who
generalized previous authors  and,
using the 1-st order formalism, suggested it   
in the closed form (without expansions)
in a Minkowski spacetime. It is also important the work by Bi\v c\'ak \cite{[3]},
where keeping quadratic approximation
on a curved Ricci flat
background, he has proved the theorem: 
with a necessity
$t^{(tot)}_{\mu\nu}$  has to contain the second derivatives of
gravitational variables.

In \cite{[4]}
the field formulation of GR was constructed on arbitrary curved
backgrounds 
with all properties of a field theory in 
a fixed spacetime. These results
obtained a development and have applications at the present time.  Thus,
in \cite{[5]}, a requiriment only of the first derivatives in
the symmetrical energy-momentum tensor and using the notions
of the work \cite{[4]} have led to a {\it new} field 
formulation of GR in a Minkowski
spacetime. Here, there is no any contradiction with the theorem
proved in the paper \cite{[3]} because in \cite{[5]} a left hand side,
unlike standard approaches, is not linear. 
In \cite{[6]}, on the basis of the work \cite{[4]} 
the total energy and angular momentum for $d+1$-dimensional
asymptotically anti-de Sitter spacetime were constructed.  
The properties of the
field approach \cite{[4]} appear under consideration in many of
concrete problems independently. For example, in \cite{[7]} and \cite{[8]}
a consideration of linear 
perturbations on a Friedmann-Robertson-Walker backgrounds leads
to the linear approximation of the full theory \cite{[4]}.
In \cite{[9a]}, in the framework of the field approach
a class of so-called `slightly bimetric'
gravitation theories was constructed, and in \cite{[9b]}
a behaviour of light cones was examined. A more full,
as we know, the  modern bibliography related
to the field approach in gravity can be also found in 
\cite{[9b]}.

Many of problems in the modern cosmology and relativistic
astrophysics are related to study of an evolution of
perturbations and conservation laws on given spacetime backgrounds.
Inflation models and discovery of accelerated expansion of
universe \cite{Chernin} initiate an interest to consideration of
perturbations on  backgrounds of de Sitter's solutions 
with the cosmological constant.
Due to more high exactness of observations it is
necessary nowadays to consider not only linear perturbations,
but next orders of them too.  It is important to construct
conservation laws for perturbations on the backgrounds,
including (anti-)de Sitter's ones (see, e.g., \cite{[6]}, 
\cite{[12]}~-~\cite{DeserTekin}).
The problems like these are also important reasons for a development
of the field formulation of GR. 
Because in  GR it is impossible to define 
in an unique way
such quantities like 
energy  there are 
many other approaches (see recent 
review \cite{Szabados}). Among them 
one is more physically consequent, it 
is the monade method for descriptions of reference systems
based on timelike word lines of observers \cite{Dehnen, Vladimirov}.
The cornestone of the construction 
is the field of a monade --- a unite  vector tangential
to a line of an observer. With its use 
the energy-momentum densities of matter as well of gravitational field 
are defined.

Return to the field approach.
In spite of all the successes 
some questions  were
unresolved up to now.  First, for constructing currents, as
usual, $ t^{(tot)\nu}_\mu$ is contracted with Killing
vectors $ \Bar \xi^\mu$ of the background \cite{[4]}: $\hat i^\nu=
\sqrt{-\Bar g} t^{(tot)\nu}_\mu \Bar \xi^\mu$.  The current is
differentially conserved 
$ \hat i^\nu{}_{;\nu} = \hat i^\nu{}_{,\nu} = 0$ if 
the covariant
differential conservation law $ t^{(tot)\nu}_\mu{}_{;\nu}
= 0$ holds. The last has a place only on flat, Ricci flat and
(anti-)de Sitter backgrounds, whereas for more complicated
backgrounds (like many of cosmological ones) $ t^{(tot)\nu}_\mu{}_{;\nu}
\neq 0$.  This fact was interpreted \cite{[4]} {\it  qualitatively only}: as
an interaction between dynamical and background systems. However,
it is necessary to give
the exact mathematical description, and we note this as a problem (A).
Second, superpotentials (antisymmetric
tensor densities, divergences of which are conserved currents)
in GR play very important role (see \cite{[10], [11]} 
and references there in). However, the construction of
superpotentials in the framework of the field formulation of GR
was not developed enough. Note it as a problem (B), concerning which
we could tell only the following.
The linear left hand side of the field equations
is  divergenceless (at least on flat backgrounds), therefore it
is already expressed through 
a superpotential.
Basing on this  Abbott and Deser \cite{[12]} have  constructed a 
superpotential with the Killing vectors on the (anti-)de Sitter
backgrounds. 

In the present paper we resolve the problems (A) and (B) and
show that they are connected. In the result, in the framework of
the field approach we construct new superpotentials and
corresponding  currents  on arbitrary curved backgrounds and
with arbitrary displacement vectors.

\section{The field formulation of GR.}
Let us present the main properties of the description  in 
\cite{[4]} 
in the 2-nd order formalism.
Consider a so-called dynamical Lagrangian \cite{[13a],[13b]}:
\be 
 \lag^{dyn} =
 \lag^E_{(dec)} -
 \hat l^{\mu\nu}
 {{\del \Bar{\lag^E}} \over {\del \Bar{\hat g^{\mu\nu}}}} -
\phi^A {{\delta \Bar {\lag^E}}\over{\delta \Bar{\Phi^A}}} -
 \Bar{\lag^E} - {1 \over {2\k }} \hat k^\alf{}_{,\alf},
\m{(1)}
\ee
the construction of which is based on the usual Lagrangian of GR
\be
\lag^E = -(2\k)^{-1} \hat R(g_{\mu\nu}) +\lag^M(\Phi^A, g_{\mu\nu})
\m{(2)}
\ee
with the metric $g_{\mu\nu}$, scalar curvature $R=g^{\alf\beta}R_{\alf\beta}$
and  the matter
variables $\Phi^A$, which are a set of arbitrary tensor densities
(not spinors).
Particular derivatives are $({}_{,\alf})$;
the hats ``$~\hat{}~$''  mean  densities of
the weight $+1$; $\delta/\delta a$ is a Lagrangain derivative;
$ \lag^E_{(dec)}$ means the Lagrangian (\ref{(2)}) after substitution 
of the decompositions
\be
 \hat g^{\mu\nu}  \equiv \Bar
 {\hat g^{\mu\nu}} + \hat l^{\mu\nu},~~~~~~~ \Phi^A  \equiv \Bar {\Phi^A}
 + \phi^A.
  \m{(3)}
 \ee
The perturbations $\hat l^{\mu\nu}$ and $\phi^A$ are 
independent gravitational and matter dynamic variables;
bars mean  background quantities which are given, 
and $\Bar  {\hat g^{\mu\nu}}$ and $\Bar {\Phi^A}$ satisfy the background
 Einstein and matter equations; indeces are shifted 
by the background metric.
We set also
\be
 \hat k^\mu =  (\Bar{ \hat g^{\mu\rho}} + \hat l^{\mu\rho})
(\Gamma^\sig_{\rho\sig} - \Bar \Gamma^\sig_{\rho\sig}) - (\Bar{\hat
 g^{\rho\sig}} + \hat
 l^{\rho\sig})  (\Gamma^\mu_{\rho\sig} - \Bar \Gamma^\mu_{\rho\sig})
 \m{(4)}
 \ee
where $\Gamma^\alf_{\mu\nu}$ are the Christoffel symbols
depending on $\hat g^{\mu\nu}$ as on the sum in ({\ref{(3)}).

It is useful to present the Lagrangian (1) as a sum of the pure
gravitational and the matter parts: $\lag^{dyn} = -({2\k})^{-1}
\lag^g + \lag^m$. 
As is seen,
the  Lagrangian (\ref{(1)})  is obtained from $\lag^E_{(dec)}$
subtracting zero's and linear's in $\hat l^{\mu\nu}$ and in
$\phi^A$ terms of the functional expansion of $\lag^E_{(dec)}$.
Zero's term is the background Lagrangian, and the linear term is
proportional to the operators of the background equations. 
The variation of $\lag^{dyn}$ with respect  to
$\hat l^{\mu\nu}$ and $\phi^A$,
some algebraic transformations, and taking into account
the background equations give
the Einstein gravitational  equations in the form
\be
 \hat
 G^L_{\mu\nu} + \hat \Phi^L_{\mu\nu} = \k\l({\hat t}^g_{\mu\nu} +
 {\hat t}^m_{\mu\nu}\r) \equiv \k{\hat t}^{(tot)}_{\mu\nu}.
  \m {(5)}
\ee
Here, the left hand side  linear in $\hat l^{\mu\nu}$ and
$\phi^A$ is expressed by
\be
 \hat G^L_{\mu\nu} \equiv {\delta
 \over {\delta\Bar{g^{\mu\nu}}}} \hat l^{\rho\sig}
 {{\delta\Bar{\hat R}}\over{\delta \Bar{\hat g^{\rho\sig}}}}\equiv
 \half \l(\hat l_{\mu\nu;\rho}{}^{;\rho} + {\Bar
 g_{\mu\nu}}{}\hat l^{\rho\sig}{}_{;\rho\sig} - {
}\hat l^{~\rho}_{\mu~;\nu\rho} - 
 \hat l^{~\rho}_{\nu~;\mu\rho}\r),
 \m{(6)}
\ee
\be
 \hat
 \Phi^L_{\mu\nu} \equiv -2\k {\delta \over
 {\delta\Bar{g^{\mu\nu}}}} \l(\hat l^{\rho\sig} 
 {{\delta\Bar{\lag^M}}\over{\delta \Bar{\hat g^{\rho\sig}}}} + \phi^A
{{\delta\Bar {\lag^M}}\over{\delta\Bar {\Phi^A}}}\r)
 \m{(7)}
\ee
 with the covariant derivatives $({}_{;\mu})$ with respect to
 $\Bar g_{\mu\nu}$. 
The right hand side of Eq. (\ref{(5)})
is the  symmetrical (metric) total energy-momentum tensor
density:
\be
 {\hat
 t}^{(tot)}_{\mu\nu} \equiv 2{{\delta\lag^{dyn}}\over{\delta
 \Bar{g^{\mu\nu}}}} \equiv 2{{\delta}\over {\delta
 \Bar{g^{\mu\nu}}}}\l(-{1\over{2\k}}\lag^g + \lag^m  \r) \equiv
 {\hat t}^g_{\mu\nu} +  {\hat t}^m_{\mu\nu}.
 \m{(8)}
\ee
Explicit expressions for $\lag^g$ and $ {\hat t}^g_{\mu\nu}$ 
follow from the Lagrangian (\ref{(1)}) and 
can be found in \cite{[4]}.

\section {Superpotentials and conserved currents.} 
To construct conservation laws 
in the field formulation of GR we use the
method developed by Katz, Bi{\v c}\'ak and Lynden-Bell (KBL) \cite{[10]}.
They are based 
on the Lagrangian:
 \be
 \lag_G=-(2\k)^{-1}\l(\hat R+{\hat k}^\mu{}_{,\mu} - \Bar{\hat R}\r)
 \m{(9)}
 \ee
depending on the physical metric $ g_{\mu\nu}$ (without decompositions)
and
the background  one $\Bar g_{\mu\nu}$; 
${\hat k}^\mu$ is also defined by Eq. (\ref{(4)}), only with
 $\hat l^{\mu\nu} = \hat g ^{\mu\nu}-\Bar{\hat g^{\mu\nu}}$.
The  identity $ \Lix \lag_G +
(\xi^\mu \lag_G)_{,\mu} \equiv 0$,
that has a place for $\lag_G$ as  a scalar density, 
is transformed 
into the main identity of the KBL approach:
\bea
&{}& \l( {{\di \lag_G} \over {\di g_{\rho\sig;\mu}}}  
g_{\rho\sig;\nu} - \lag_G \delta^\mu_\nu\r)\xi^\nu + 
2\l({{\di \lag_G} \over {\di  g_{\nu\tau;\mu}}} 
g_{\nu(\tau}\delta^\rho_{\lam)} \Bar
 g^{\lam\sig}\r)\xi_{[\sig,\rho]} \nonumber\\ 
 &{}& +~ 2\l( {{\delta
{\lag_G}} \over {\delta g_{\rho\sig}}} { g_{\rho(\sig}}
\delta^\mu_{\nu)}  + {{\delta {\lag_G}} \over {\delta \Bar
g_{\rho\sig}}} { \Bar g_{\rho(\sig}} \delta^\mu_{\nu)} \r)\xi^\nu+
\hat \zeta^\mu \equiv  \hat I^{\mu\nu}{}_{;\nu}
\equiv \hat I^{\mu\nu}{}_{,\nu}.
 \m{(10)}
\eea
Here, the Lie derivative, say, of a vector $\psi^\alf$ with respect to  
$\xi^\alf$ is defined  as 
$
 \Lix \psi^\alf \equiv -
\xi^\beta  \psi^\alf{}_{;\beta} + \xi^\alf{}_{;\beta}\psi^\alf.
$ 
In Eq. (\ref{(10)}),
(i) the first term 
$(\hat t^\mu_\nu  +(2\k)^{-1}\hat
l^{\rho\sig}\bar R_{\rho\sig} \delta^\mu_\nu)\xi^\nu$
includes the canonical energy-momentum tensor density $\hat t^\mu_\nu$
for the free gravitational field; (ii) the second term is
presented by the spin tensor density
$\hat\sig^{\mu\rho\sig}$ contracted with
$\xi_{[\sig;\rho]}$;
(iii) the third term is 
${\k}^{-1}({\hat G^\mu_\nu }-
{\Bar{\hat G^\mu_\nu} })\xi^\nu$ with the Einstein tensor
$G^\mu_\nu$;
(iv) the fourth term is equal to zero if $\xi^\nu$ is a Killing vector
of the background. Antisymmetric tensor density $\hat I^{\mu\nu}$
on the right hand side of (\ref{(10)}) is 
the KBL superpotential generalizing 
the well known  Freud  superpotential \cite{[14]}
on arbitaray curved backgrounds.
Explicitly all the expressions can be found in \cite{[10]}.
Using the dynamic ${\hat G^\mu_\nu }=\k{{\hat T^\mu_\nu} }$ 
and  background $\Bar{\hat G^\mu_\nu }=\k{\Bar{\hat T^\mu_\nu} }$  
Einstein equations,
KBL transform the identity 
(\ref{(10)}) into a `weak' conservation law 
for the  current $\hat I^\mu$
($\di_\mu \hat I^\mu= 0$):
\be
\l[\hat t^\mu_\nu + ({\hat T^\mu_\nu }-{\Bar{\hat T^\mu_\nu} }) 
+(2\k)^{-1}\hat
 l^{\rho\sig}\bar R_{\rho\sig} \delta^\mu_\nu
 \r]\xi^\nu +\hat\sig^{\mu\rho\sig}\xi_{[\sig,\rho]}
+ \hat \zeta^\mu \equiv
\hat I^\mu = 
\hat I^{\mu\nu}{}_{,\nu}.
\m{(11)}
\ee
To apply the KBL technique in the field approach we 
use $ \hat g^{\mu\nu} - \Bar {\hat
g^{\mu\nu}}$  instead of 
$\hat l^{\mu\nu}$. Then, the gravitational part $-(2\k)^{-1} \lag^g$ of
(\ref{(1)}) depending on the first derivatives of $\hat l^{\mu\nu}$ only
 goes to $\lag^{(2)}$ and is expressed over
the KBL Lagrangian (9) as $\lag^{(2)}  \equiv
 \lag_G - \lag^{(1)}$ with $ \lag^{(1)} = 
 -(2\k)^{-1} (\hat g^{\mu\nu} - \Bar {\hat
 g}^{\mu\nu}) \Bar R_{\mu\nu}$.
After that we transform the identity 
$
 \Lix \lag^{(2)} +  (\xi^\mu \lag^{(2)})_{,\mu}
\equiv 0
$
into
\bea 
&{}& \l({{\di \lag_{G}} \over {\di 
g_{\rho\sig;\mu}}}  g_{\rho\sig;\nu} - \lag_G\delta^\mu_\nu +
\lag^{(1)} \delta^\mu_\nu - 2{{\delta  \lag^{(1)}} \over {\delta g_{\rho\sig}}}
g_{\rho(\sig} \delta^\mu_{\nu)}
\r) \xi^\nu - {\hat
{\cal M}}^{(2)\mu\rho}_\lam \Bar g^{\sig\lam}\xi_{\sig;\rho}\nonumber\\
 &{}& +~ 
2 \l({{\delta \lag_G} \over {\delta g_{\rho\sig}}}
g_{\rho(\sig}\delta^\mu_{\nu)}  + {{\delta \lag^{(2)}}
\over {\delta \bar g_{\rho\sig}}} \Bar g_{\rho(\sig}\delta^\mu_{\nu)}
\r) \xi^\nu \equiv
- \left({\hat{\cal M}}^{(2)\mu\nu}_\lam \xi^\lam\right)_{,\nu}.
 \m {(12)}
\eea
Here,
 \be 
 \hat{\cal M}^{(2)\mu\nu}_\lam \equiv -2\l({{\di
\lag^{(2)}} \over {\di g_{\rho\sig;\mu}}} {
g_{\rho(\sig}} \delta^\nu_{\lam)} + {{\di \lag^{(2)}} \over
{\Bar g_{\rho\sig,\nu}}} {\Bar g_{\rho(\sig}}
\delta^\nu_{\lam)}\r)
 \m{(13)}
\ee
is antisymmetric in $\mu$ and $\nu$
and  expressed over
the spin term presented in Eqs. (\ref{(10)}) - (\ref{(11)}) as
\be
 {\hat {\cal M}}^{(2)\mu\nu}_\lam\Bar g^{\lam\rho} =
 \hat\sig^{\rho[\mu\nu]}+
 \hat\sig^{\mu[\rho\nu]}-\hat\sig^{\nu[\rho\mu]}.
 \m{(14)}
 \ee
Noting that $\hat G^{L}_{\mu\nu}$ in (\ref{(6)})
can be thought as depending on $\hat g^{\mu\nu}$ as well 
on $\hat l^{\mu\nu}$,
it is important to find that in Eq. (\ref{(12)}):
 $$ 2{{\delta  \lag^{(2)}} \over
{\delta \Bar g_{\rho\sig}}} { \Bar g}_{\rho(\sig}
\delta^\mu_{\nu)}  \equiv - {1 \over {\k}}\hat G^{L\mu}_{\nu}.
 $$
Subtracting Eq. (\ref{(12)}) from Eq. (\ref{(10)}), 
replacing $\hat g^{\mu\nu}$ by $\hat l^{\mu\nu}$
in correspondence with (\ref{(3)}) 
we obtain the  identity:
\be
 {1 \over {\k}} \hat G^{L\mu}_{\nu}\xi^\nu + {1 \over
 \k} \hat l^{\mu\lam} \Bar
 R_{\lam\nu}\xi^\nu + \hat \zeta^\mu_{(*)} \equiv  \hat
 I^{\mu\nu}_{(*)}{}_{;\nu} \equiv  \hat
 I^{\mu\nu}_{(*),\nu}
 \m {(15)}
\ee
with the new superpotential 
$ \hat I^{\mu\nu}_{(*)} =
\hat I^{\mu\nu} +{\hat {\cal M}}^{(2)\mu\nu}_\lam\Bar g^{\lam\rho}\xi_\rho$.
The quantity  $ \hat \zeta^\mu_{(*)}  = \hat \zeta^\mu + 
{\hat {\cal M}}^{(2)\mu\rho}_\lam\Bar g^{\lam\sig}\xi_{(\sig;\rho)}$
is equal to zero on 
Killing vectors of a background, 
like $\hat \zeta^\mu$ in (\ref{(10)}) and (\ref{(11)}).

To obtain weak conservation laws one has to substitute
Einstein's equations (\ref{(5)})  into the identity
(\ref{(15)}).  Note that the quantity $\hat \Phi^{L}_{\mu\nu}$
defined in  Eq.  (\ref{(7)}) can be excluded from Eq. (\ref{(5)}).
Indeed, substituting (\ref{(1)}) and (\ref{(2)}) into the definition
(\ref{(8)}) one finds that the expression (\ref{(7)}) appears
also on the right hand side of equations (\ref{(5)}), and thus
they are rewritten as
\be
\hat G^L_{\mu\nu} = \k\l( {\hat t}^g_{\mu\nu} +  
\delta {\hat t}^M_{\mu\nu} \r)
 \m {(16)}
\ee
with
\be 
\delta{\hat t}^M_{\mu\nu} \equiv 2 {{\delta} \over {\delta
\Bar{g^{\mu\nu}}}}
\l[\lag^M \l(\Bar \Phi^A +
\phi^A,~  \Bar{\hat g^{\mu\nu}} + \hat l^{\mu\nu}\r) - 
\Bar{\lag^M}\r].
 \m{(17)}
\ee
Substituting  Eq. (\ref{(16)}) into Eq. (\ref{(15)}) we
obtain:
 \be
 \l({\hat t}_{\nu}^{g\mu} +\delta {\hat
t}_{\nu}^{M\mu}  + { \k}^{-1} \hat
l^{\mu\lam} \Bar R_{\lam\nu}\r)\xi^\nu + \hat \zeta^\mu_{(*)} 
\equiv \hat {\cal T}^\mu_{(*)\nu}\xi^\nu + \hat
\zeta^\mu_{(*)}
\equiv \hat I^{\mu}_{(*)} =
 \hat I^{\mu\nu}_{(*)}{}_{,\nu}.
 \m {(18)}
\ee
It gives a conserved law
 $
 \di_\mu \hat I^{\mu}_{(*)} = 0 $
for the  current $\hat I^{\mu}_{(*)}$ that takes a place
(i) on arbitrary curved backgrounds including
all cosmological solutions; (ii) for 
arbitrary displacement
vectors $\xi^\mu$, not only for the Killing ones.
(iii) 
We present also  the explicit term 
$\k^{-1}\hat l^{\mu\lam} \Bar R_{\lam\nu}$ 
(included into $\hat {\cal T}^\mu_{(*)\nu}$),
which describes the
interaction with a background.
(iv) Constructing
Eq. (\ref{(18)}) we suggested a new superpotential,  
the explicit form of which
after calculations in (\ref{(15)}) with (\ref{(13)})
and the KBL superpotential is given by
 \be
  \hat I^{\mu\nu}_{(*)} =
 {\k}^{-1} \hat
 l^{\rho[\mu}\xi^{\nu]}{}_{;\rho} +  {\k}^{-1}
 \l(\Bar g^{\rho[\mu} \hat l^{\nu]\sig}-
 \Bar g^{\sig[\mu} \hat l^{\nu]\rho}\r)_{;\sig}
 \xi_\rho,
 \m{(19)}
\ee
where the coefficient at $\xi_\rho$ is a covariantized Papapetrou 
superpotential \cite{[15]}.

Thus, the new conservation law in the form of the 
expression (\ref{(18)}) 
resolves the problems (A) and (B) annonced in the first section.
The question arises: why we have a success with Eq. (\ref{(18)})? First,
instead of $\hat t^{m}_{\mu\nu}$ in Eq.
(\ref{(5)}) we use the modified matter energy-momentum density
$ \delta{\hat t}_{\mu\nu}^{M}$ in Eq. (\ref{(16)}) with the 
definition (\ref{(17)}), that does not follow by the
usual way from the  Lagrangian (\ref{(1)}). Second,
due to the term $\k^{-1}\hat l^{\mu\lam} \Bar R_{\lam\nu}$ the new
energy-momentum tensor density $\hat {\cal T}^{\mu\nu}_{(*)}$ is
{\it non-symmetrical} on general complicated backgrounds,
whereas before only the
{\it symmetrical}
$\hat t^{(tot)}_{\mu\nu}$ was used.

Let us do remarks. 1) Currents and superpotentials in
\cite{[11]} obtained with applying the Belinfante method to the
canonical system developed in \cite{[10]}.  They exactly
coincide with the conserved quantities given here in
(\ref{(18)}) and (\ref{(19)}).  Thus the Belinfante procedure
becomes `a bridge' connecting two approaches for constructing
conservation laws: {\it canonical} \cite{[10]} and {\it
symmetrical} \cite{[4]} ones.  2) The field formulation of GR
can be constructed with the use of the decomposition $g_a = \Bar
g_a + h_a$ of any metrical variable from the set $g_a \in
g_{\mu\nu},\, g^{\mu\nu},\, \sqrt{-g}g_{\mu\nu},\,
\ldots$ \cite{[13b]}. Then  field equations, currents and
superpotentials acquire the forms respectively of  (\ref{(5)}),
(\ref{(18)}) and  (\ref{(19)}) with the replacement of $\hat
l^{\mu\nu}$ by $\hat l^{\mu\nu}_a = h_a(\di \Bar {\hat
g}^{\mu\nu}/\di\Bar g_a)$.  The use of different definitions for
$\hat l^{\mu\nu}_a$ leads to an ambiguity in the definition of
energy-momentum tensor (as firstly was noted in
\cite{[16]}) and superpotentials beginning from the
2-nd order in perturbations. On the other hand, in
\cite{[10],[11]} the construction of the currents and
superpotentials does not depend on a choice of a dynamical
variable from the set $g_a$. Therefore, results \cite{[11]}
coinciding only with (\ref{(18)}) and (\ref{(19)}), resolve this
ambiguity in favour of the decomposition (\ref{(3)}).

\medskip
\noindent {\bf Acknowledgments.} The author very thanks 
Joseph Katz for useful discussions and fruitful
recommendations, he
is very grateful to Ji\v r\'i Bi\v c\'ak, Stephen Lau and
L\'aszl\'o Szabados
for explanations of their works and helpful conversations.

\ed